\documentclass[reprint,showpacs,preprintnumbers,amsmath,amssymb,prc,floatfix]{revtex4-1}
\usepackage{float}
\usepackage{color}
\usepackage{graphicx}
\usepackage{dcolumn}
\usepackage{bm}
\usepackage{xcolor}
\usepackage{comment}
\usepackage[colorlinks,citecolor=blue,linkcolor=red,anchorcolor=blue,filecolor=blue,urlcolor=blue]{hyperref}
\usepackage{multirow}
\usepackage{threeparttable} 
\usepackage{adjustbox} 
\usepackage{color}
\usepackage{longtable}
\usepackage{csquotes}
\begin{document}
\title{Medium dependent relativistic NN potential: Application to the fusion dynamics} 
\author{M. Bhuyan$^{1,2,3}$} 
\email{bunuphy@um.edu.my}
\author{Shilpa Rana$^4$} 
\email{srana60\_phd19@thapar.edu}
\author{Nishu Jain$^4$}
\email{nishujain1003@gmail.com}
\author{Raj Kumar$^4$}
\email{rajkumar@thapar.edu}
\author{S. K. Patra$^{5}$}
\author{B. V. Carlson$^{6}$}
\bigskip
\affiliation{$^1$Department of Physics, Faculty of Science, University of Malaya, Kuala Lumpur 50603, Malaysia}
\affiliation{$^2$Institute of  Research and Development, Duy Tan University, Da Nang 550000, Vietnam}  
\affiliation{$^3$Faculty of Natural Science, Duy Tan University, Da Nang 550000, Vietnam}
\affiliation{$^4$School of Physics and Materials Science, Thapar Institute of Engineering and Technology, Patiala 147004, India}
\affiliation{$^5$Institute of Physics, Sachivalaya Marg, Sainik School, Bhubaneswar 751005, India}
\affiliation{$^6$Institute Tecnologico de Aeroneutica, Sao Jose dos Campos, Sao Paulo, 1222900, Brazil}
\date{\today}
\bigskip 
\begin{abstract}
\noindent
In-medium effects are introduced in the microscopic description of the effective nucleon-nucleon (NN) interaction potential entitled DDR3Y in terms of the density-dependent nucleon-meson couplings within the Relativistic-Hartree-Bogoliubov (RHB) approach. The nuclear densities of the interacting target and projectile nuclei and NN potentials are obtained for non-linear NL3$^*$ and TM1 parameter sets within the relativistic mean-field approach and density-dependent DDME1 and DDME2 parameter sets within the Relativistic-Hartree-Bogoliubov (RHB) formalism. The DDR3Y NN potential and the densities are used to obtain the nuclear potential by adopting the double folding approach. This nuclear potential is further used to probe the fusion dynamics within the $\ell-$summed Wong model for a few {\it even-even} systems leading to the formation of light, heavy and superheavy nuclei. The calculations are also performed for the relativistic R3Y, density- dependent and independent M3Y interaction potentials for the comparison. We observed that the DDR3Y NN potential gives a better overlap with the experimental data as compared to non-relativistic M3Y and DDM3Y NN potentials. From the comparison of R3Y and DDR3Y interactions, it is manifested that the inclusion of in-medium effects in terms of density-dependent nucleon-meson couplings raises the fusion barrier and consequently decreases the fusion and/or capture cross-section. Moreover, the nuclear densities, as well as the relativistic R3Y NN potential obtained for the NL3$^*$ parameter set, are observed to give a comparatively better fit to the experimental data.
\end{abstract}

\maketitle
\section{INTRODUCTION}\label{intro}
The interaction barrier generated by two colliding nuclei is of the essence in understanding complex nuclear reaction dynamics. The formulae for the long-range repulsive Coulomb and centrifugal potentials formed between two interacting heavy ions are straightforward, whereas the short-range attractive nuclear potential evaluation is ambiguous. Despite numerous theoretical efforts, the understanding of the nuclear interaction in the total interaction potential is still fuzzy \cite{mont17,toub17,canto20,das98,raj20}. The double-folding model \cite{satc79} is one of the widely used techniques that turned out to give a satisfactory description of the real part of the nucleus-nucleus as well as $\alpha-$nucleus optical potentials. In the double folding model, the nuclear optical potential is obtained by integrating the densities of the interacting nuclei over an effective nucleon-nucleon (NN) interaction potential. The widely adopted choices of the effective NN interaction potential are the M3Y (Michigan 3 Yukawa) interactions which were developed to fit the G-matrix elements of Reid  \cite{bert77} and Paris \cite{anan83} NN potentials on an oscillator basis. Later, the density dependence was included in the original M3Y interactions to account for the higher-order exchange effects and the Pauli blocking exchange effects. Moreover, the density-independent M3Y interactions were also observed to fail in saturating cold nuclear matter (NM) \cite{bert77,anan83}. Consequently, numerous density-dependent M3Y NN interactions were developed with density-dependent parameters fitted to reproduce the saturation properties of NM.

Parallel to the M3Y potential, recently in Refs. \cite{sing12,sahu14,lahi16}, the relativistic NN potential was derived within the well-established relativistic mean-field (RMF) formalism and entitled R3Y NN potential. Furthermore, this relativistic R3Y NN potential was employed to study the cluster radioactivity \cite{sing12,sahu14,lahi16,joe22} and fusion dynamics \cite{bhuy18,bhuy20,rana21,rajprc22,ranaprc22} of various even-even, even-odd, and odd-odd reactions leading to the synthesis of heavy and superheavy nuclei. From these works, it is concluded that the results obtained from the R3Y NN interactions provide a relatively better overlap with the experimental data. It is worth mentioning that the R3Y NN interaction is obtained for the linear and non-linear relativistic parameter sets. Here our main aim is to introduce the medium effects in the R3Y NN-potential in terms of density-dependent nucleon-meson couplings within the Relativistic-Hartree-Bogoliubov Approach, which will be analogous to the density-dependent M3Y (DDM3Y) potential.
 
Unlike the non-relativistic M3Y NN interactions, the relativistic R3Y NN potential is given in terms of meson masses and coupling constants. The non-linear coupling constants in the relativistic R3Y NN potential include effectively the medium effects \cite{boguta77}. Furthermore, the other reliable method to introduce medium dependence in the relativistic R3Y NN interaction potential is to include density-dependent coupling constants, which is different from the DDM3Y NN potential. For example, the medium effect was introduced in M3Y potential through a weighted function of the density \cite{bert77,anan83}, however for the R3Y NN potential, the medium effect will be included through nucleon-meson coupling constants within the Relativistic-Hartree-Bogoliubov (RHB) approach \cite{niks02,lala05}. The density-dependent DDME1 and DDME2 parameter sets are used in the present study to obtain the relativistic R3Y NN potential entitled DDR3Y potential. Furthermore, the DDR3Y potential, along with the densities from the RHB approach, is used to estimate the fusion dynamics of various heavy-ion reactions. Here, we have chosen six reaction systems namely, $^{16}$O+$^{40}$Ca, $^{40}$Ca+$^{58}$Ni, $^{40}$Ca+$^{90}$Zr, $^{16}$O+$^{144}$Sm, $^{16}$O+$^{208}$Pb and $^{48}$Ca+$^{208}$Pb forming light, heavy and superheavy nuclei to examine the application of the DDR3Y NN potential in the description of fusion. The results of the newly introduced DDR3Y NN potential obtained for the DDME1 \cite{niks02} and DDME2 \cite{lala05} parameter sets within the RHB approach are also compared with the R3Y NN potential obtained for the non-linear NL3$^*$ \cite{lala09} and TM1 \cite{bodmer91} RMF parameter sets. Moreover, the traditional Reid M3Y and the density-dependent M3Y (DDM3Y) \cite{satc79,kobo82,khoa94} are also considered for comparison. The fusion and/or capture cross-section is obtained within the $\ell-$summed Wong model \cite{kuma09,wong73} and results are also compared with the available experimental data \cite{vigdor79,morton94,bour14,timm98,das97,prok08}.

The paper is organized as follows: The details of the theoretical formalism adopted in the present analysis are explained in section \ref{theory}. The results obtained from the theoretical calculations are discussed in section \ref{results}. In Section \ref{summary}, a summary and conclusions of the work are presented.
     
\section{Theoretical formalism}
\label{theory} 
A large number of nucleons are involved in the nuclear fusion of two interacting heavy ions. The interaction potential formed between two colliding heavy ions plays a vital role in elucidating the complex fusion mechanism. The total interaction potential can be written as the sum of three terms, 
\begin{eqnarray}
V_{T}^{\ell}(R)=V_C(R)+ V_{\ell}(R)+ V_n (R).
\label{vtot}
\end{eqnarray}
Here, R is the separation distance between the interacting nuclei. The terms $V_C(R)=Z_pZ_te^2/R$ and $V_{\ell}(R)=\frac{\hbar^2\ell(\ell+1)}{2\mu R^2}$ are the repulsive Coulomb and centrifugal potentials, respectively. $Z_p$ and $Z_t$  symbolize the charge numbers of projectile and target nuclei, and $\mu$ is the reduced mass. The last term in Eq. (\ref{vtot}) denotes the short-range and attractive nuclear potential, which is calculated within the double-folding approach \cite{satc79} as,
\begin{eqnarray}
V_{n}(\vec{R}) & = &\int\boldsymbol{\rho}_{p}(\vec{r}_p)\boldsymbol{\rho}_{t}(\vec{r}_t)V_{eff}
\left(\boldsymbol{\rho}, r{\equiv} |\vec{r}_p-\vec{r}_t +\vec{R}| \right) \nonumber \\
&& d^{3}r_pd^{3}r_t. 
\label{fold}
\end{eqnarray}
Here, bold symbols $\boldsymbol{\rho}_p$ and $\boldsymbol{\rho}_t$ denote the total nuclear density (i.e. sum of proton and neutron densities) distributions of the interacting projectile and target nuclei, respectively. $V_{eff}(\boldsymbol{\rho},r)$ is the effective nucleon-nucleon (NN) interaction potential. As discussed above, the density-independent M3Y (Michigan 3 Yukawa) NN interactions are widely-adopted, and the density-dependent M3Y (DDM3Y) NN interactions have also been developed to include the in-medium effects. The relativistic R3Y NN potential comparable to the M3Y has also been recently derived from the relativistic mean-field (RMF) formalism. The present analysis aims to obtain the density-dependent R3Y (DDR3Y) NN potential from the relativistic-Hartree-Bogoliubov (RHB) model. The M3Y, DDM3Y, R3Y, and DDR3Y NN potentials are discussed in detail in the following subsections.

\subsection{M3Y and DDM3Y Nucleon-Nucleon Interaction Potentials}
The explicit medium dependence in the original M3Y interaction was introduced \cite{satc79,kobo82,khoa94,khoa95} via multiplying it by a density-dependent weight factor $F(\boldsymbol{\rho})$, 
\begin{eqnarray}
V_{eff}^{M3Y}(\boldsymbol{\rho},r)= F(\boldsymbol{\rho})V_{eff}^{M3Y}(r). 
\label{ddm3y}
\end{eqnarray}
Here, $V_{eff}^{M3Y}(r)$ is the radial dependent M3Y NN interaction. In the present analysis, we have considered the widely-adopted Reid M3Y \cite{bert77} interaction which is written as the sum of three Yukawa terms as,
\begin{eqnarray}
V_{eff}^{M3Y}(r)=7999 \frac{e^{-4r}}{4r}-2140\frac{e^{-2.5r}}{2.5r}+J_{00}\delta(r). 
\label{m3y}
\end{eqnarray}
Here, $J_{00}(E)\delta(r)$ is the long range one-pion exchange potential (OPEP). The different versions of the density dependent $F(\boldsymbol{\rho})$ factor are developed in the literature with parameters fitted to reproduce the saturation properties of the nuclear matter. Here, we have considered the BDM3Y-type \cite{khoa95}, which is written as,
\begin{eqnarray}
F(\boldsymbol{\rho})=C[1-\alpha\boldsymbol{\rho}^{\beta}].
\label{ddf}
\end{eqnarray}
As mentioned above, the parameters C, $\alpha$ and $\beta$ are adjusted to match the nuclear matter saturation properties. Here, we have adopted the BDM3Y1 (C=1.2253, $\alpha=1.5124$ $fm^3$ and $\beta=1.0$) version \cite{khoa95} of the density dependent Reid M3Y NN interaction as it yields a nuclear incompressibility value $K=232$ MeV, which is comparable to the ones given by NL3$^*$ ($K=258.8$ MeV) \cite{lala09}, TM1 ($K=281$ MeV) \cite{bodmer91}, DDME1 ($K=244.5$ MeV) \cite{niks02} and DDME2 ($K=250.89$ MeV) \cite{lala05} parameter sets considered here. The density $\boldsymbol{\rho}$ in Eq. (\ref{ddf}) is taken as sum of the projectile and target densities at the midpoint of the nucleon-nucleon separation distance. This procedure is known as the frozen density approximation (FDA) \cite{satc79,kobo82,khoa94,khoa95,soub01,khoa07} and is widely-adopted in the folding model.

\subsection{R3Y and DDR3Y Nucleon-Nucleon Interaction Potentials}
\begin{figure}
\centering
\includegraphics[scale=0.32]{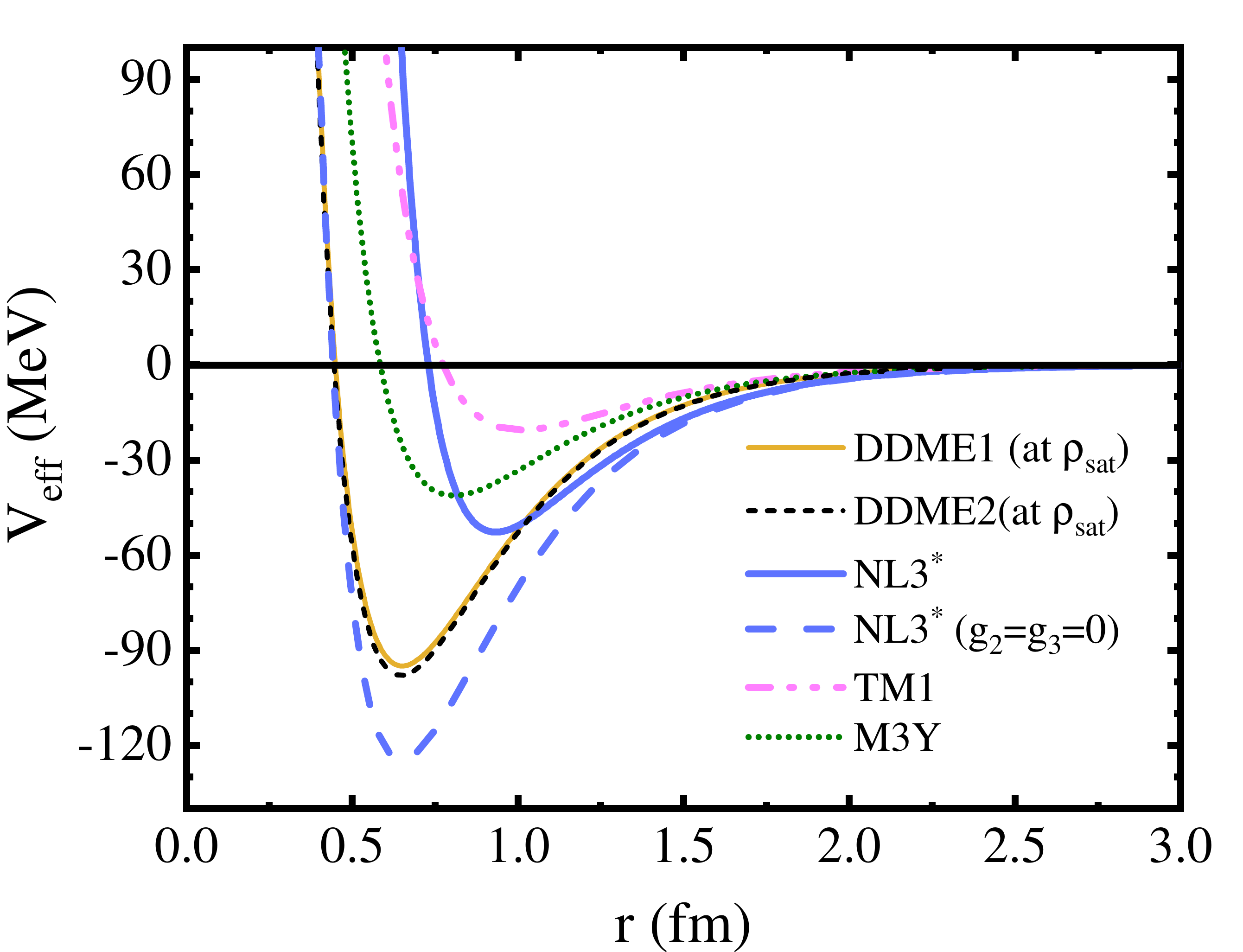}
\caption{(Color online) The R3Y nucleon-nucleon (NN) potential for DDME1 (solid orange line), DDME2 (dashed black line), NL3$^*$ (thick solid blue line) and TM1 (thick dash double dotted magenta line) parameter sets are compared with the well-known M3Y (dotted green line) potential. For sake of presentation in 2D, the NN potentials for DDME1 and DDME2 are calculated using the couplings at saturation density ($\rho_{sat}=0.152$ $fm^{-3}$ \cite{niks02,lala05}). The thick dashed line signifies the R3Y NN potential obtained without non-linear coupling terms. See text for details.}
\label{fig1}
\end{figure}

The relativistic effective nucleon-nucleon R3Y interaction \cite{sing12,sahu14,lahi16} is derived from the self-consistent relativistic mean-field (RMF) formalism. The RMF models have emerged to be very reliable in the description of structural properties of the finite nuclei, not only in the $\beta$-stable regions but also in the regions of extreme isospin asymmetry lying close to the drip lines. A phenomenological Lagrangian density  \cite{vret05,tani20,afan96,agbe15,meng16,ring96,lala09,sing12,sahu14,lahi16,meng06,bhuy18,bhuy20,rana21,rajprc22,ranaprc22,joe22,bodmer91} describing the the nucleon-mesons many body system can be written as,
\begin{eqnarray}
{\cal L}&=&\overline{\psi}\{i\gamma^{\mu}\partial_{\mu}-M\}\psi +{\frac12}\partial^{\mu}\sigma
\partial_{\mu}\sigma \nonumber -{\frac12}m_{\sigma}^{2}\sigma^{2}\\
&& -{\frac13}g_{2}\sigma^{3} -{\frac14}g_{3}\sigma^{4}
-g_{\sigma}\overline{\psi}\psi\sigma \nonumber -{\frac14}\Omega^{\mu\nu}\Omega_{\mu\nu}\\
&& +{\frac12}m_{\omega}^{2}\omega^{\mu}\omega_{\mu}+{\frac14}\xi_3(\omega^{\mu}\omega_{\mu})^2
-g_{\omega}\overline\psi\gamma^{\mu}\psi\omega_{\mu} \nonumber \\
&&-{\frac14}\vec{B}^{\mu\nu}.\vec{B}_{\mu\nu}+\frac{1}{2}m_{\rho}^2
\vec{\rho}^{\mu}.\vec{\rho}_{\mu} -g_{\rho}\overline{\psi}\gamma^{\mu}
\vec{\tau}\psi\cdot\vec{\rho}^{\mu}\nonumber \\
&&-{\frac14}F^{\mu\nu}F_{\mu\nu}-e\overline{\psi} \gamma^{\mu}
\frac{\left(1-\tau_{3}\right)}{2}\psi A_{\mu}.
\label{lag}
\end{eqnarray}
Here, $m_\sigma$, $m_\omega$ and $m_\rho$ are the masses of the corresponding $\sigma$, $\omega$, and $\rho$ mesons, which intermediate the interaction between the nucleons of mass {$M$} denoted by the Dirac spinor $\psi$. $g_\sigma$, $g_\omega$ and $g_\rho$ are the nucleon-meson coupling constants and $g_2$, $g_3$ and $\xi_3$ denote the non-linear meson self interaction constants. The quantities $\tau$ and $\tau_3$ in Eq. (\ref{lag}) denote the isospin and its third component, respectively. $\Omega^{\mu\nu}$, $\vec B^{\mu\nu}$ and $F^{\mu\nu}$ symbolize the field tensors for $\omega$, $\rho$ and photons, respectively and are written as,
\begin{eqnarray}
F_{\mu\nu} =\partial_{\mu} A_{\nu}-\partial_{\nu} A_{\mu}  \\
\Omega_{\mu\nu} = \partial_{\mu} \omega_{\nu} - \partial_{\nu} \omega_{\mu}
\end{eqnarray}
and
\begin{eqnarray}
\vec{B}_{\mu \nu}=\partial_{\mu} \vec{\rho}_{\nu} -\partial_{\nu} \vec{\rho}_{\mu}.
\end{eqnarray}
Here, $A_\mu$ is the electromagnetic field. The equations of motions for the Dirac nucleons and mesons can be derived using the Euler-Lagrange equations in the mean-field approximation and are written as, 
\begin{eqnarray}
&& \Bigl(-i\alpha.\nabla+\beta(M+g_{\sigma}\sigma)+g_{\omega}\omega+g_{\rho}{\tau}_3{\rho}_3 \Bigr){\psi} = {\epsilon}{\psi}, \nonumber \\
&& \left( -\nabla^{2}+m_{\sigma}^{2}\right) \sigma(r)=-g_{\sigma}{\rho}_s(r)-g_2\sigma^2 (r) - g_3 \sigma^3 (r),\nonumber  \\ 
&& \left( -\nabla^{2}+m_{\omega}^{2}\right) \omega(r)=g_{\omega}{\rho}(r)-\xi_3\omega^3(r),\nonumber   \\  
&& \left( -\nabla^{2}+m_{\rho}^{2}\right) \rho(r)=g_{\rho}{\rho}_3(r).
\label{field}
\end{eqnarray} 
The relativistic effective nucleon-nucleon interaction entitled R3Y potential \cite{sing12,sahu14,lahi16} is derived from the mean-field equations for the mesons under the limit of one meson exchange. The relativistic R3Y NN potential is obtained for linear and non-linear RMF parameter sets and has been applied successfully in exploring the various nuclear phenomenon. In the recent studies, \cite{sing12,sahu14,lahi16,joe22,bhuy18,bhuy20,rana21,rajprc22,ranaprc22}, the Reid M3Y, as well as the R3Y NN potentials, are used and these studies show that the R3Y NN potential obtained for non-linear RMF parameter sets give comparatively better overlap with the available experimental data. As discussed above, the explicit density dependence was introduced in the M3Y NN potential to account for the in-medium effects, which also results in a better description of infinite nuclear matter properties. In this direction, it is crucial and interesting to introduce medium dependence in the relativistic R3Y NN potential. Here within relativistic mean-field formalism, it is not necessary to multiply a weighted density function with the NN potential to make it medium dependent, as shown for the M3Y potential in Eq. (\ref{ddm3y}). The non-linear meson self-interactions terms in the R3Y NN potential obtained for non-linear RMF parameter sets effectively include the medium effects \cite{boguta77}. The alternate method to include medium effects in the relativistic NN potential (R3Y) is to consider the density-dependent parametrization within the Relativistic-Hartree-Bogoliubov (RHB) approach, where the nucleon-meson couplings are medium dependent and are defined as  \cite{niks02,lala05,type99,fuchs95,hofm01}, 
\begin{eqnarray}
g_i(\boldsymbol{\rho})=g_i(\boldsymbol{\rho}_{sat})f_i(x)|_{i=\sigma,\omega},
\label{dd1}
\end{eqnarray}
where
\begin{eqnarray}
f_i(x)=a_i\frac{1+b_i(x+d_i)^2}{1+c_i(x+d_i)^2}
\label{dd2}
\end{eqnarray}
and
\begin{eqnarray}
g_\rho(\boldsymbol{\rho})=g_\rho(\boldsymbol{\rho}_{sat})exp[-a_\rho(x-1)].
\label{dd3}
\end{eqnarray}
Here, $x=\boldsymbol{\rho} / \boldsymbol{\rho}_{sat}$, with $\boldsymbol{\rho}_{sat}$ being the baryon density of symmetric nuclear matter at saturation. The five constraints- $f_i(1)=1$, $f''_i(0)=0$, and $f''_\sigma(1)=f''_\omega(1)$ reduce the number of independent parameters in Eq. (\ref{dd2}) from eight to three. The independent parameters (the meson mass and coupling parameters) of the RHB formalism are obtained to fit the ground state properties of finite nuclei as well as the properties of symmetric and asymmetric nuclear matter. In the present analysis, we have adopted the well-known DDME1 \cite{niks02} and DDME2 parameter sets \cite{lala05} to study the fusion mechanism of various reactions. The density-dependent R3Y (DDR3Y) NN potential ($V_{eff}^{R3Y}(r,\boldsymbol{\rho}_p, \boldsymbol{\rho}_t)$) in terms of density-dependent meson-nucleon coupling constants defined above can be written as,
\begin{eqnarray}
V_{eff}^{R3Y}(r,\boldsymbol{\rho}_p, \boldsymbol{\rho}_t)&=&\sum_{i=\omega,\rho}\frac{{g_{i}(\boldsymbol{\rho}_p) g_{i}(\boldsymbol{\rho}_t)}}{4{\pi}}\frac{e^{-m_{i}r}}{r} \nonumber \\
& - &\frac{{g_{\sigma}(\boldsymbol{\rho}_p) g_{\sigma}(\boldsymbol{\rho}_t)}}{4{\pi}}
\frac{e^{-m_{\sigma}r}}{r} + J_{00}\delta(r). 
\label{ddr3y}
\end{eqnarray}
Here, $m_\sigma$, $m_\omega$ and $m_\rho$ are the masses of the corresponding $\sigma$, $\omega$, and $\rho$ mesons, which intermediate the interaction between the nucleons. $g_\sigma$, $g_\omega$ and $g_\rho$ are the nucleon-meson coupling constants and $J_{00}(E)\delta(r)$ is the long-range one-pion exchange potential (OPEP). The expression for the DDR3Y in Eq. (\ref{ddr3y}) is identical in form to that of the R3Y NN potential used in previous studies of Refs. \cite{sing12,sahu14,lahi16,joe22,bhuy18,bhuy20,rana21,ranaprc22}. Here in the DDR3Y, the nucleon-meson coupling constants are density-dependent, while they are constant in the case of the R3Y NN potential. The relativistic DDR3Y NN potential is obtained for the DDME1 (solid orange line) and DDME2 (dashed black line) within the Relativistic-Hartree-Bogoliubov approach. The  R3Y NN potential is also calculated for non-linear NL3$^*$ (thick solid blue line) and TM1 (thick dash double dotted magenta line) parameter sets within the relativistic mean-field formalism and non-relativistic M3Y NN potential (dotted green line) for comparison. The results for the NN  potential are shown as a function of the nucleon separation (r) in Fig. \ref{fig1}. It is worth mentioning that, the R3Y NN potential for the DDME1 and DDME2 parameter sets is plotted at the saturation density ($\rho_{sat}=0.152$ $fm^{-3}$ for DDME1 \cite{niks02} and DDME2 \cite{lala05}). More details on the relativistic parametrizations used in the present analysis can be found in Refs. \cite{niks02,lala05,lala09,bodmer91}.

As the DDME1 and DDME2 parameter sets do not contain any self-interacting non-linear terms in the meson field, we have also given the R3Y potential for the NL3$^*$ parameter set without the non-linear meson self-interaction terms (thick dashed blue line) for the sake of comparison. It can be observed from Fig. \ref{fig1} that the R3Y NN potential obtained for the NL3$^*$ parameter set without the non-linear terms shows the deepest pocket, followed by the DDME2 and DDME1 parameter sets at saturation density. This indicates that the inclusion of the non-linear meson interaction terms gives a repulsive core to the NN potentials, which is essential to reproduce the saturation properties of infinite nuclear matter. In the relativistic-Hartree-Bogoliubov approach, the medium-dependent nucleon-meson vertices are introduced instead of non-linear self-interaction terms, and more details can be found in Refs. \cite{niks02,lala05}. Following this, the influence of the density-dependent R3Y NN potential (see Eq. (\ref{ddr3y})) within the RHB approach will be taken into account through nuclear potential within the double folding model.

\begin{figure}
\centering
\includegraphics[scale=0.32]{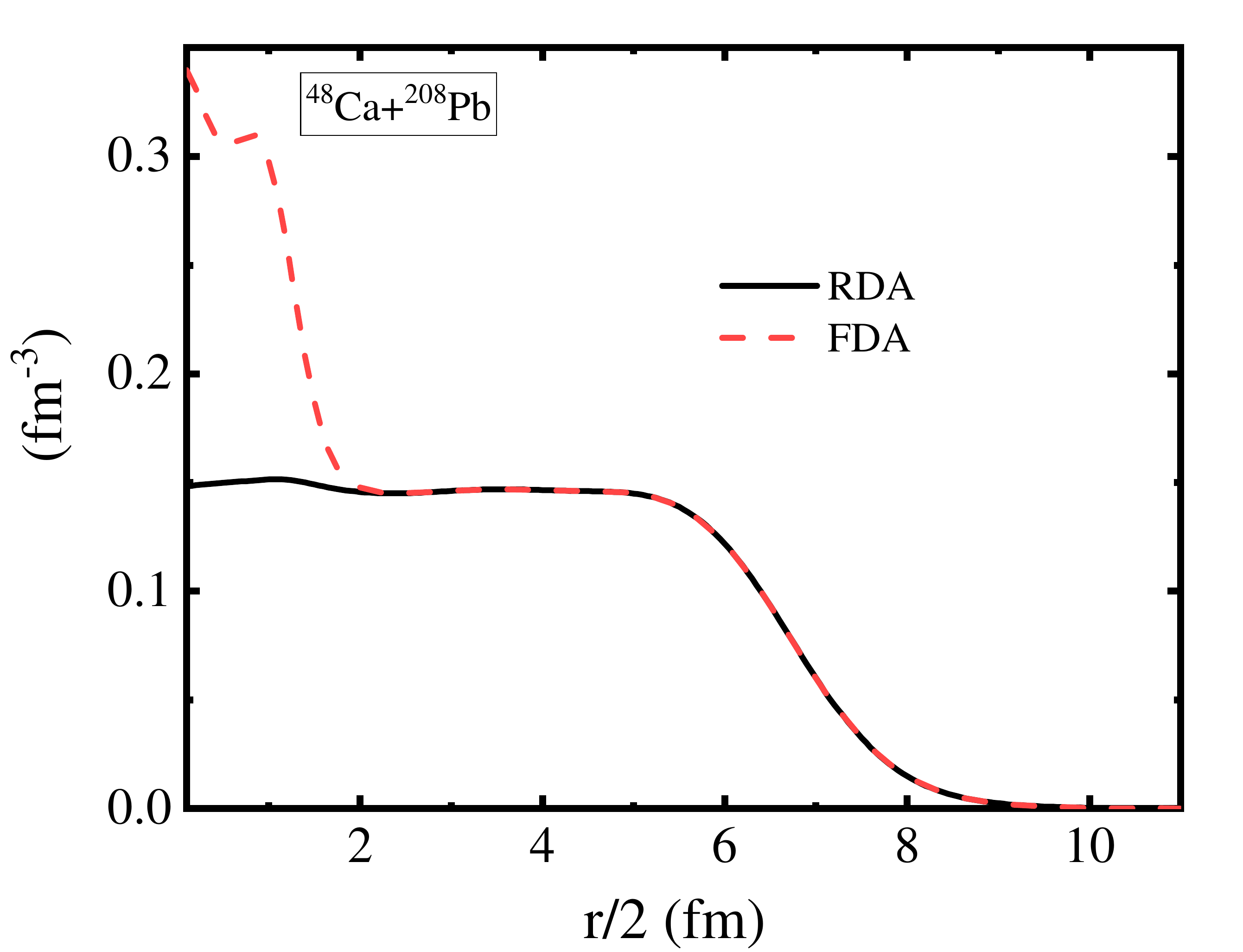}
\caption{(Color online) The nuclear density ($\boldsymbol{\rho}$) at mid-point of inter-nucleon separation ($r/2$) obtained within the frozen density approximation (dashed red line) and relaxed density approximation (solid black line) for $^{48}$Ca+$^{208}$Pb system.}
\label{fig2}
\end{figure}

It is worth noting here that the DDR3Y NN potential (see Eq. (\ref{ddr3y})) depends upon both the projectile and target densities, unlike the DDM3Y (see Eq. \ref{ddm3y}), which is given in terms of the density at the midpoint of the nucleon separation. This is because the relativistic DDR3Y NN potential is obtained microscopically within the RHB approach and the terms $[|g_{i}(\boldsymbol{\rho}_p) g_{i}(\boldsymbol{\rho}_t)|_{i=\sigma,\omega,\rho}]$ account for the meson exchange between the nucleons of the projectile and target nuclei. Thus in the DDR3Y, the density dependence is introduced microscopically in terms of meson-exchange, as shown explicitly in Eq. (\ref{ddr3y}). As mentioned above, in case of the DDM3Y NN potential, the density dependence is introduced through a weight function $F(\boldsymbol{\rho})$ and $\boldsymbol{\rho}$ is obtained within the frozen density approximation (FDA) \cite{satc79,kobo82,khoa94}. There is another parallel method that exists to introduce the density for a composite system, the so-called relaxed density approximation (RDA) \cite{denisov13,denisov15}. In RDA, the nuclear fusion around and below the barrier is assumed to be a slow process that allows the relaxation of proton and neutron densities \cite{denisov13,denisov15}. The FDA and RDA are widely adopted in the double folding approach for the M3Y potential, as shown in Eq. (\ref{ddf}) and the Skyrme Energy Density functional, respectively. Here in Fig. \ref{fig2}, we have shown a comparison of the DDME2 density ($\boldsymbol{\rho}$) at the midpoint of nucleon separation \cite{khoa94} for an illustrative case of $^{48}$Ca+$^{208}$Pb system calculated within the FDA (dashed red line) and RDA (solid black line). It can be observed from the Fig. \ref{fig2} that $\boldsymbol{\rho}$ obtained within the FDA for lower values of $r/2$ is much higher than the nuclear matter saturation density ($\boldsymbol{\rho}_{sat}=0.152$ $fm^{-3}$). This is because, in the FDA, the density at a fixed point of space is given by the sum of the nucleon densities of both nuclei \cite{satc79,kobo82,khoa94,khoa95,soub01,khoa07}, which exceeds $\boldsymbol{\rho}_{sat} = 0.152$ $fm^{-3}$ at the smaller nucleon distances. This problem can be resolved by adopting the RDA in which the density at any separation does not surpass the equilibrium density of nuclear matter \cite{denisov13,denisov15}.

\begin{figure}
\centering
\includegraphics[scale=0.32]{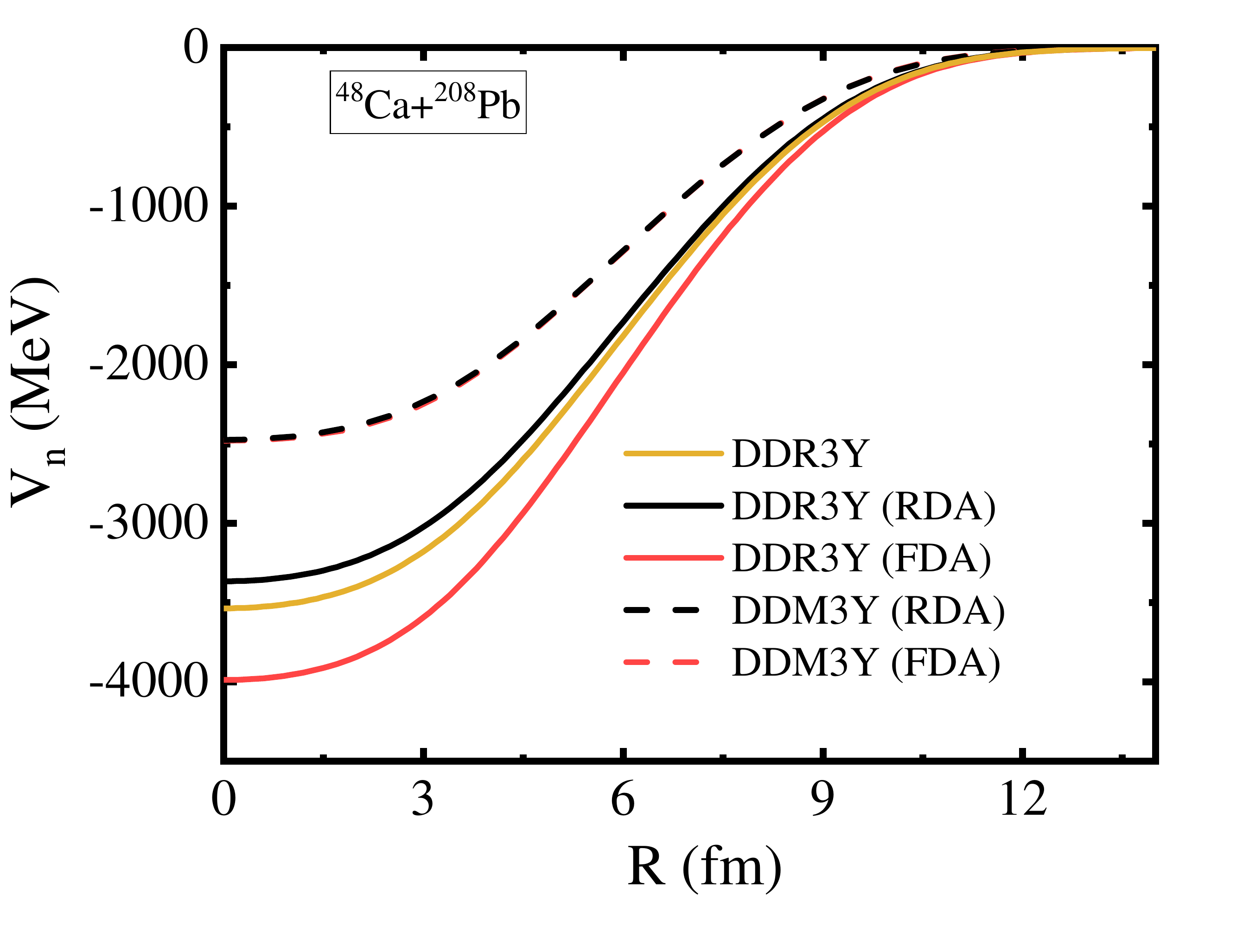}
\caption{(Color online) Nuclear potential ($V_n$) calculated within frozen density approximation (red) and relaxed density approximation (black) using DDM3Y (dashed) and DDR3Y (solid) NN potentials for $^{48}$Ca+$^{208}$Pb system.}
\label{fig3}
\end{figure}

To assess the validity of the FDA and RDA for evaluating the DDM3Y, we have calculated the nuclear potential using both approximations within the double folding approach. For the sake of comparison, we have also evaluated the DDR3Y NN potential using both these approximations (i.e. replacing $\boldsymbol{\rho}_p$ and  $\boldsymbol{\rho}_t$ in Eq. (\ref{ddr3y}) by $\boldsymbol{\rho}$). Fig. \ref{fig3} shows the nuclear potential calculated using the DDM3Y (dashed lines) and DDR3Y (solid lines) within the FDA (red) and RDA (black) for the case of $^{48}$Ca+$^{208}$Pb system. The solid orange line shows the nuclear potential calculated using the DDR3Y given in Eq. (\ref{ddr3y}). It is worth mentioning here that to obtain the nuclear potential using Eq. (\ref{ddr3y}), neither of the frozen or relaxed density approximations is needed. It can be noticed clearly from Fig. \ref{fig3} that there is a remarkable difference between the nuclear potentials obtained within the FDA and RDA for the DDR3Y NN potential. Comparing the results obtained from the FDA and RDA with those from Eq. (\ref{ddr3y}), it is observed that the FDA yields much attractive nuclear potential, whereas the RDA provides slightly repulsive nuclear potential. Moreover, the results of the RDA are closer to those obtained using Eq. (\ref{ddr3y}) than the FDA. In the case of the DDM3Y, the difference between the nuclear potential obtained within the FDA and RDA is almost negligible. This is because, in the case of the DDM3Y, the density dependence enters through a weight function $F (\boldsymbol{\rho})$ whereas for the DDR3Y the nucleon-meson couplings of $\sigma-$, $\omega-$ and $\rho-$mesons are density-dependent (see Fig. 1 in \cite{niks02}). Thus the FDA approximation, which gives $\boldsymbol{\rho}>> \boldsymbol{\rho}_{sat}$ at smaller $r$ is inappropriate for obtaining the nuclear potential using the DDR3Y NN potential. The RDA, however, gives results close to those obtained using Eq. (\ref{ddr3y}), but still, there is a noticeable difference, especially at smaller separation distance (R). 
However, both the RDA as well as the FDA give similar results in the case of the DDM3Y. Following all these observations, we have adopted the traditional FDA \cite{satc79,kobo82,khoa94} for calculating the nuclear potential using the DDM3Y. To obtain the nuclear potential using the DDR3Y NN potential, no approximation to the density is used, and the medium dependence is introduced in terms of both projectile and target density-dependent nucleon-meson couplings. More explicitly, the medium dependence in DDR3Y is introduced directly in the coupling constants. 

The nuclear potentials obtained by folding different density distributions and NN interactions are further employed to study the fusion of various systems within the extended $\ell-$summed Wong model \cite{bhuy18,bhuy20,rana21,rajprc22,ranaprc22,kuma09}. This $\ell-$summed Wong model \cite {kuma09} is the refined version of the simple Wong formula \cite{wong73} and accounts for the  actual modifications entering the fusion barrier due to its angular momentum dependence. The cross-section in terms of $\ell$ -partial wave is written as,
\begin{eqnarray}
\sigma(E_{c.m.})=\frac{\pi}{k^{2}} \sum_{\ell=0}^{\ell_{max}}(2\ell+1)P_{\ell}(E_{c.m}).
\label{crs}
\end{eqnarray}
Here, $k=\sqrt{\frac{2 \mu E_{c.m.}}{\hbar^{2}}}$ and $E_{c.m.}$ denotes the energy of target-projectile system in the center of mass frame. In present study, $\ell_{max}$ values are extracted from the sharp cut-off model \cite{beck81} for above barrier energies and extrapolated for below barrier energies. The quantity $P_{\ell}$ symbolize the transmission coefficient. Using the Hill-Wheeler \cite{hill53} approximation of symmetric parabolic barrier, $P_{\ell}$ can be written in terms of barrier height ($V_B^\ell$) and curvature ($\hbar \omega_{\ell}$) as,
\begin{eqnarray}
P_{\ell}=\Bigg[1+exp\bigg(\frac{2 \pi (V_{B}^{\ell}-E_{c.m.})}{\hbar \omega_{\ell}}\bigg)\Bigg]^{-1}. 
\label{hw}
\end{eqnarray}
 The barrier characteristics such as the barrier height ($V_B^\ell$) and barrier position ($R_B^\ell$) can be evaluated easily once the total interaction potential (see Eq. (\ref{vtot})) is defined, i.e.,
 \begin{eqnarray}
 \frac {dV_{T}^{\ell}}{dR} \bigg |_{R=R_{B}^{\ell}}=0.
 \label{vb1}
 \end{eqnarray}
 \begin{eqnarray}
 \frac{d^2V_{T}^{\ell}}{dR^2}\bigg |_{R=R_{B}^{\ell}}\le 0.
 \label{vb2}
 \end{eqnarray}
Further, the barrier curvature ($\hbar\omega_\ell$) can also be evaluated at $R = R_{B}^{\ell}$ corresponding to the barrier height $V_{B}^{\ell}$ as,
\begin{eqnarray}
\hbar \omega_{\ell}=\hbar [|d^2V_{T}^{\ell}(R)/dR^2|_{R=R_{B}^{\ell}}/\mu]^{\frac{1}{2}}.
\label{vb3}
\end{eqnarray}  
These barrier characteristics of the total interaction potential are obtained using M3Y, DDM3Y, R3Y, and DDR3Y NN interaction potential described in detail above.
\begin{figure}[ht]
\centering
\includegraphics[scale=0.32]{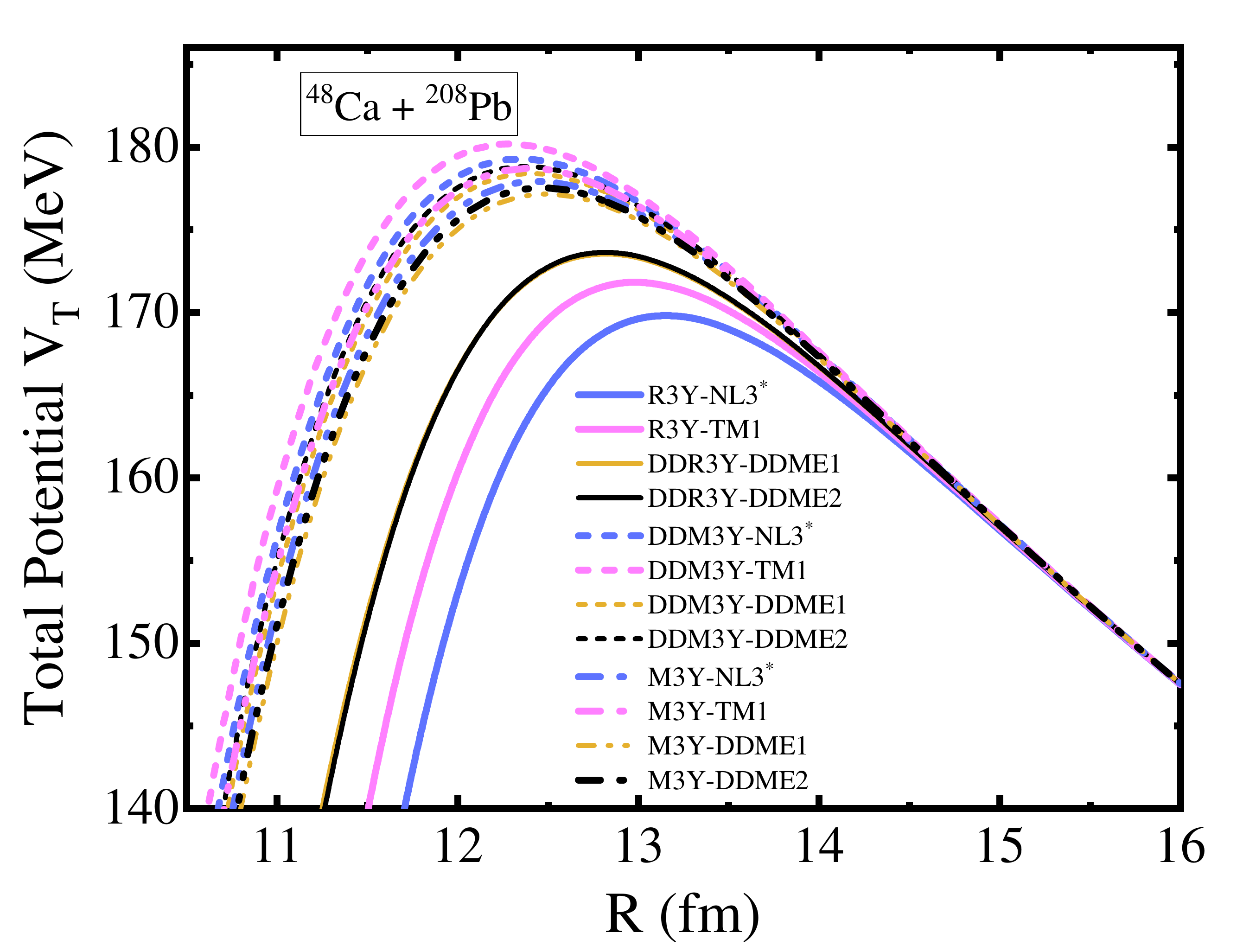}
\caption{(Color online) The total interaction potential $V_T$ (MeV) at $\ell =0\hbar$ as a function of radial separation R for $^{48}$Ca+$^{208}$Pb system calculated using the M3Y (dash double dotted lines), DDM3Y (dashed lines), R3Y (thick solid lines) and DDR3Y (solid lines) NN potentials. The blue (thick dark grey), magenta (thick light grey), orange (light grey) and black lines are for NL3$^*$, TM1, DDME1 and DDME2 parameters sets, respectively. See text for details.}
\label{fig4}
\end{figure}
 \section{RESULTS AND DISCUSSION}
\label{results}
\begin{table*}
\centering
\caption{The positions $R_B$ (in fm) and heights $V_B$ (in MeV) of the fusion barriers obtained using R3Y, DDR3Y, M3Y, and DDM3Y NN potentials folded with nuclear densities obtained within RHB and RMF approaches for all the reactions under study. }
\begin{tabular}{ccccccccccccc}
\hline
Reaction          & \multicolumn{2}{c}{$^{16}$O+$^{40}$Ca} & \multicolumn{2}{c}{$^{40}$Ca+$^{58}$Ni} & \multicolumn{2}{c}{ $^{40}$Ca+$^{90}$Zr} & \multicolumn{2}{c}{$^{16}$O+$^{144}$Sm} & \multicolumn{2}{c}{$^{16}$O+$^{208}$Pb} & \multicolumn{2}{c}{$^{48}$Ca+$^{208}$Pb} \\ \hline
Nuclear Potential & $R_B$           & $V_B$            & $R_B$            & $V_B$            & $R_B$            & $V_B$            & $R_B$            & $V_B$             & $R_B$           & $V_B$            & $R_B$           & $V_B$            \\ \hline
R3Y-NL3$^*$         & 9.4          & 22.85         & 10.5          & 72.36         & 11.2         & 96.83          & 11.3          & 59.59          & 12.2          & 73.17         & 13.2         & 169.81         \\
R3Y-TM1           & 9.2          & 23.26         & 10.3          & 73.45         & 11.0         & 98.20          & 11.1          & 60.46          & 12.0          & 74.11         & 13.0         & 171.84         \\
DDR3Y-DDME1       & 9.2          & 23.35         & 10.2          & 73.91         & 10.9         & 99.18          & 11.0          & 60.88          & 11.9          & 74.80         & 12.8         & 173.55         \\
DDR3Y-DDME2       & 9.2          & 23.41         & 10.2          & 74.03         & 10.9         & 99.28          & 10.9          & 60.99          & 11.9          & 74.90         & 12.8         & 173.64         \\
DDM3Y-NL3$^*$       & 8.8          & 24.28         & 9.8           & 76.70         & 10.4         & 102.73         & 10.7          & 62.66          & 11.5          & 76.89         & 12.3         & 179.28         \\
DDM3Y-TM1         & 8.7          & 24.53         & 9.7           & 77.29         & 10.4         & 103.45         & 10.6          & 63.12          & 11.4          & 77.36         & 12.3         & 180.20         \\
DDM3Y-DDME1       & 8.9          & 24.06         & 9.8           & 76.13         & 10.5         & 102.12         & 10.7          & 62.24          & 11.6          & 76.43         & 12.4         & 178.41         \\
DDM3Y-DDME2       & 8.8          & 24.17         & 9.8           & 76.41         & 10.5         & 102.41         & 10.7          & 62.44          & 11.5          & 76.65         & 12.4         & 178.83         \\
M3Y-NL3$^*$          & 8.9          & 24.01         & 9.8           & 76.34         & 10.5         & 101.77         & 10.7          & 62.20          & 11.6          & 76.31         & 12.4         & 177.93         \\
M3Y-TM1           & 8.8          & 24.23         & 9.7           & 76.84         & 10.5         & 102.40         & 10.7          & 62.64          & 11.5          & 76.74         & 12.4         & 178.72         \\
M3Y-DDME1         & 9.0          & 23.80         & 9.9           & 75.63         & 10.6         & 101.22         & 10.8          & 61.81          & 11.7          & 75.88         & 12.5         & 177.17         \\
M3Y-DDME2         & 8.9          & 23.90         & 9.9           & 75.87         & 10.6         & 101.49         & 10.8          & 62.00          & 11.6          & 76.09         & 12.5         & 177.54         \\ \hline
\end{tabular}
\label{tab1}
\end{table*}
\begin{figure*}[t]
\centering
\includegraphics[scale=0.22]{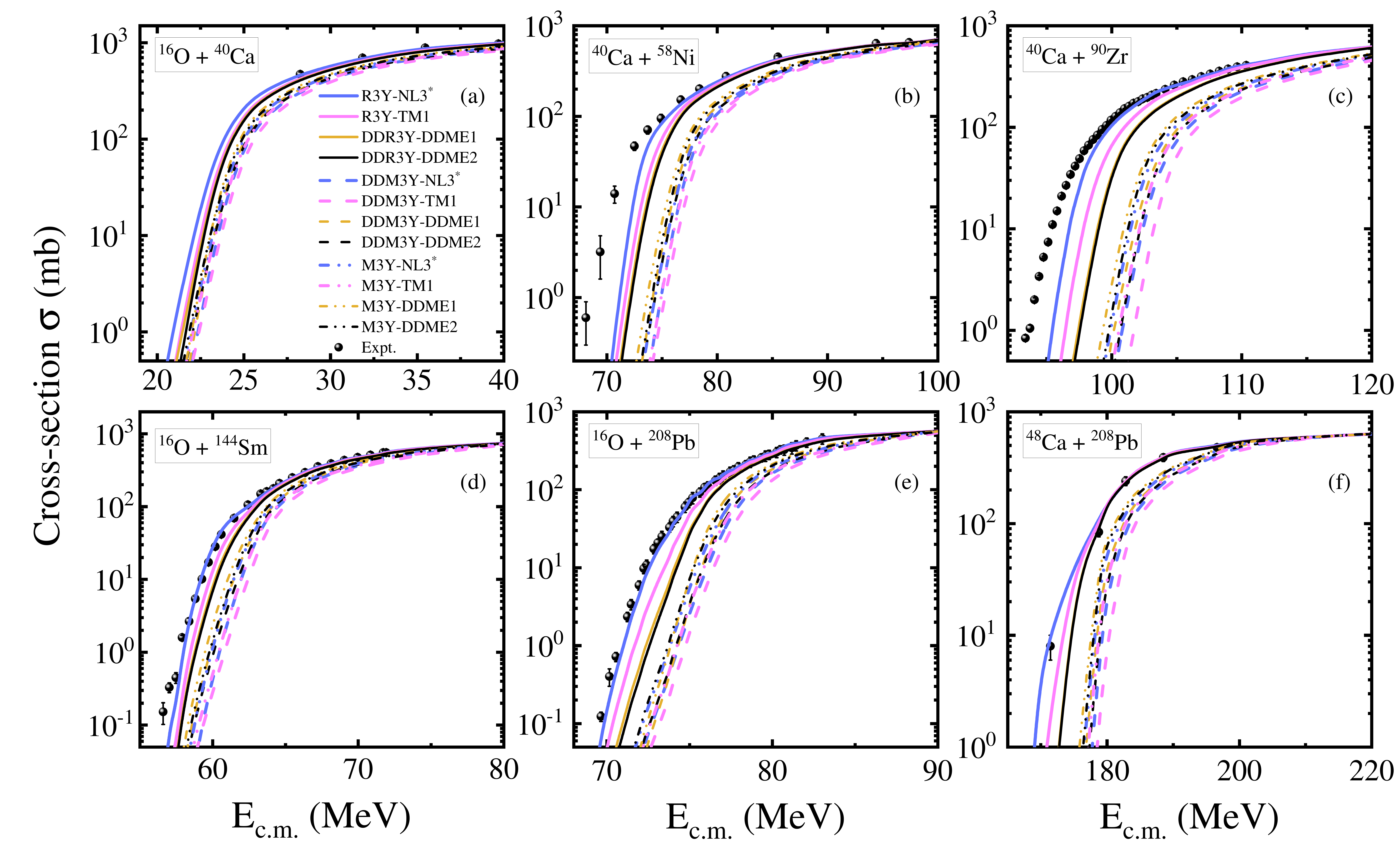}
\caption{(Color online) The cross-section $\sigma$ ($mb$) for all target-projectile combinations considered in the present study calculated using the M3Y (dash double dotted lines), DDM3Y (dashed lines), R3Y (thick solid lines) and DDR3Y (solid lines) NN potentials. The blue (thick dark grey), magenta (thick light grey), orange (light grey) and black lines are for NL3$^*$, TM1, DDME1 and DDME2 parameters sets, respectively. See text for details.}
\label{fig5}
\end{figure*}
This section aims to assess the application of medium-dependent relativistic DDR3Y NN potential (see Eq. (\ref{ddr3y})) in the description of nuclear fusion dynamics. For this a detailed analysis of the barrier characteristics and fusion and/or capture cross-section for six heavy-ion reactions namely  $^{16}$O+$^{40}$Ca, $^{40}$Ca+$^{58}$Ni, $^{40}$Ca+$^{90}$Zr, $^{16}$O+$^{144}$Sm, $^{16}$O+$^{208}$Pb and $^{48}$Ca+$^{208}$Pb is carried out. These even-even, spherical and/or nearly spherical \cite{moller16} target-projectile systems lead to the formation of compound nuclei with $N/Z$ ratios varying from 1 (isospin symmetric) to 1.5 (isospin asymmetric). As mentioned above, the nuclear potential formed between two colliding heavy ions provides a key to elucidating the fusion mechanism, by introducing in-medium effects in the microscopic description of the relativistic NN interaction potential. The density-dependent microscopic DDR3Y NN potential is obtained within the well-established relativistic-Hartree-Bogoliubov (RHB) approach for the DDME1 \cite{niks02}, and DDME2 \cite{lala05} parameter set. It is worth mentioning that unlike \cite{yahya21}, here, the medium effects in the DDR3Y are introduced microscopically in terms of density-dependent couplings of the $\sigma-$, $\omega-$, and $\rho-$mesons. Also, unlike the DDM3Y, no approximation is used to introduce the density dependence in the relativistic DDR3Y NN potential. One can observe from Fig.  \ref{fig1} that all the NN potentials have similar forms but with different depths. The non-linear TM1 parameter set provides the most repulsive core, whereas the NL3$^*$ parameter without non-linear meson interaction terms results in the most attractive core of the R3Y NN potential. It is worth noting here that the R3Y NN potential for DDME1  and DDME2 parameter sets is plotted at the saturation density in Fig. \ref{fig1}, whereas the actual density-dependent R3Y NN potential given by Eq. (\ref{ddr3y}) is used in the double folding approach to obtain the nuclear potential.

The nuclear densities for the target and projectile nuclei considered are also obtained within the RHB approach for the DDME1 and DDME2 parameter sets and the RMF formalism for the NL3$^*$ and TM1 parameter sets. Folding these densities with the R3Y, DDR3Y, DDM3Y, and M3Y effective NN potentials, we get 12 sets of nuclear potentials, namely R3Y-NL3$^*$, R3Y-TM1, DDR3Y-DDME1, DDR3Y-DDME2, DDM3Y-NL3$^*$, DDM3Y-TM1, DDM3Y-DDME1, DDM3Y-DDME2, M3Y-NL3$^*$, M3Y-TM1, M3Y-DDME1, and M3Y-DDME2 for each reaction. As discussed in section \ref{theory}, the frozen density approximation results in densities ($\boldsymbol{\rho}$) much higher than the nuclear matter saturation density ($\boldsymbol{\rho}_{sat}$) at small separation (see Fig. \ref{fig2} for the case of $^{48}$Ca+$^{208}$Pb system for the DDME2 set). This problem is resolved when the nuclear density at the mid-point of separation is used within the relaxed density approximation (RDA). However, it can be observed from Fig. \ref{fig3}, that the nuclear potential obtained for DDM3Y within the FDA and RDA almost overlap. In the case of the DDR3Y, neither the FDA nor RDA is required, as it incorporates the meson-exchange effect microscopically within the RHB approach between the projectile and target nuclei (see Eq. (\ref{ddr3y})). Thus both target and projectile densities are used to introduce medium-dependence in the DDR3Y NN potential and consequently to calculate the DDR3Y-DDME1 and DDR3Y-DDME2, whereas the FDA is used to evaluate DDM3Y-NL3$^*$, DDM3Y-TM1, DDM3Y-DDME1, and DDM3Y-DDME2 nuclear potentials. Further investigation of Fig. \ref{fig3} shows that the relativistic DDR3Y NN potential gives a more attractive nuclear interaction potential as compared to the DDM3Y NN potential. For a more comprehensive study, the repulsive Coulomb potential is added to all the 12 sets of nuclear potentials mentioned above.

Fig. \ref{fig4} shows the barrier regions of the s-wave ($\ell=0 \hbar$) total interaction potentials obtained for $^{48}$Ca+$^{208}$Pb reaction using all 12 combinations of nuclear densities and effective NN potentials. The positions $R_B$ (in fm) and heights $V_B$ (in MeV) of the fusion barrier for the six reactions under study are listed in Table \ref{tab1}. Here, DDR3Y-DDME1 and DDR3Y-DDME2 signify that both the effective NN potentials as well as the nuclear densities are obtained within the RHB approach for the density-dependent DDME1 and DDME2 parameter sets, respectively. The BDM3Y1 version of the density-dependent Reid NN interactions (denoted as DDM3Y) is considered here since it gives a nuclear incompressibility (K) value comparable to that given by the RMF parameter sets considered in the present study. It can be observed from Fig. \ref{fig4}, as well as Table \ref{tab1}, that the R3Y-NL3$^*$ nuclear potential gives the lowest barrier, whereas the DDM3Y-TM1 provides the highest fusion barrier for the considered heavy-ion reactions. From the comparison of barrier heights obtained for the R3Y and newly introduced DDR3Y, as well as those obtained for the M3Y and DDM3Y NN potentials, it is observed that the inclusion of density dependence in the effective NN potential increases the fusion barrier because the in-medium effects result in a more repulsive nuclear potential (see Fig. \ref{fig3}). However, the difference between the barrier characteristics obtained for the DDR3Y and R3Y is more prominent than that for the DDM3Y and M3Y NN potentials. This is because, in the DDR3Y, the medium-dependence is introduced microscopically in terms of the density-dependent meson-nucleon couplings. Further, comparing the results given by different nuclear density distributions folded with the same M3Y NN potentials, it is noted that densities obtained for the DDME1 and TM1 parameters sets give the lowest and highest barrier, respectively. Also, the total potentials obtained within the DDME1 and DDME2 almost overlap in the barrier region, with the DDME2 giving a slightly higher barrier. The characteristics of the total interaction potentials obtained for different NN potentials and nuclear densities are further used to calculate the fusion probability.

The $\ell-$summed Wong model equipped with the relativistic-Hartree-Bogoliubov and relativistic mean-field approaches is used to evaluate the fusion and/or capture cross-section for the six reactions under study. The $\ell-$values are obtained using the sharp cut-off model \cite{beck81} at above barrier center of mass energies and are extrapolated for the below-barrier region. The fusion and/or capture cross-section $\sigma$ (mb) as function of the center of mass-energy ($E_{c.m.}$) is shown in Fig. \ref{fig5}, calculated using 12 different nuclear potentials listed in Table \ref{tab1}. It can be noted from Fig. \ref{fig5} that among the R3Y, DDR3Y, M3Y, and DDM3Y, the highest cross-section is obtained for the R3Y NN potential, whereas the DDM3Y yields the lowest cross-section for all the systems. From the comparison of results obtained for the NL3$^*$, TM1, DDME1, and DDME2 densities folded with the same NN potential (M3Y and DDM3Y), it is observed that the DDME1 and TM1 densities give the highest and lowest cross-sections, respectively. On comparing the cross-section obtained for newly developed DDR3Y with that obtained for non-linear R3Y NN potential, it is observed that the fusion and/or capture cross-section decreases on introducing the medium-dependence through the density-dependent nucleon-meson couplings.

The experimental data are taken from Refs. \cite{vigdor79,morton94,bour14,timm98,das97,prok08} for comparison at below and above barrier energies as shown in Fig. \ref{fig5}. The R3Y NN potential, as well as the nuclear densities obtained for the NL3$^*$ parameter set, are observed to give a comparatively better fit to the experimental data than the other sets of nuclear potentials. As discussed above, the inclusion of in-medium effects in the DDR3Y NN potential within the RHB approach decreases the cross-section which results in the under-estimation of the experimental data. Both the non-relativistic M3Y and DDM3Y NN potentials are also observed to underestimate the fusion and/or capture cross-section for all the systems under study. For the $^{40}$Ca+$^{58}$Ni (Fig. \ref{fig5}(b)) and $^{40}$Ca+$^{90}$Zr (Fig. \ref{fig5}(d)) systems, the R3Y-NL3$^*$ is also observed to underestimate the cross-section at below barrier energies. The difference between the cross-section obtained using different nuclear potentials is smallest for the $^{16}$O+$^{40}$Ca reaction, involving comparatively lighter and doubly magic target and projectile nuclei. Moreover, the difference between cross-sections obtained using different nuclear potentials decreases progressively as we move towards the higher center of mass energies. This is because, at above barrier energy regions, the effects of nuclear structure are diminished considerably, and only the angular momentum effect persists \cite{bhuy18,bhuy20}. Also, the difference between cross-sections obtained for different RMF and RHB parameter sets increases with the increase in the mass of the compound nucleus formed in the reaction. This indicates that a relevant choice of nuclear potential becomes more and more important as we move towards the exotic regions of the nuclear chart.

\section{SUMMARY AND CONCLUSIONS}
\label{summary}
The density dependence is introduced in the description of the relativistic R3Y effective NN potential. The R3Y NN potential is obtained in terms of density-dependent nucleon-meson couplings within the framework of the relativistic-Hartree-Bogoliubov (RHB) approach for the well-known DDME1 and DDME2 parameter sets. This newly developed effective NN potential is entitled DDR3Y NN potential and is further employed to obtain the nuclear potential within the double folding approach and also the fusion and/or capture cross-section within the $\ell-$summed Wong model. The microscopic R3Y NN potential obtained within relativistic mean-field (RMF) formalism for non-linear NL3$^*$ and TM1 parameter sets, as well as the non-relativistic M3Y and DDM3Y NN potentials, are also considered for the comparison. The frozen density approximation (FDA) is used to calculate the DDM3Y NN potential, whereas no such approximation is needed to obtain the microscopic DDR3Y NN potential as it is obtained in terms of both projectile and target density-dependent nucleon-meson couplings within the RHB approach. The comparison of the fusion barrier characteristics and fusion and/or capture cross-sections obtained within different forms of the NN potential (R3Y, DDR3Y, M3Y and DDM3Y) is carried out for six reactions namely $^{16}$O+$^{40}$Ca, $^{40}$Ca+$^{58}$Ni, $^{40}$Ca+$^{90}$Zr, $^{16}$O+$^{144}$Sm, $^{16}$O+$^{208}$Pb and $^{48}$Ca+$^{208}$Pb leading to the formation of light, heavy and superheavy compound nuclei.

From the comparison of barrier characteristics and cross-sections obtained for the R3Y and newly introduced DDR3Y as well as those obtained for the M3Y and DDM3Y NN potentials, it is observed that the inclusion of density dependence in the effective NN potential increases the fusion barrier, which consequently decreases the fusion and/or capture cross-section. However, the difference between results obtained for the R3Y and DDR3Y is considerably more prominent than those obtained for the M3Y and DDM3Y NN potentials. This is because the medium-dependence is introduced in the DDR3Y via density-dependent meson-nucleon couplings, unlike in the DDM3Y, where the density-dependent is introduced through a weight function. Further, from the comparison of cross-sections obtained using different nuclear potentials with the experimental data, it is noticed that the DDR3Y NN potential underestimates the fusion and/or capture cross-section for all the considered reactions. However, the match between the experimental and theoretical cross-section is better for the DDR3Y NN potentials than for the DDM3Y NN potentials. The relativistic R3Y NN potential and nuclear densities obtained for the NL3$^*$ parameter set are observed to provide a comparatively better fit to the experimental data.  Moreover, the difference between the different RMF and RHB parameter sets increases with the increase in the mass number of the compound nuclei. Thus, a more systematic study involving more degrees of freedom and nuclear reactions forming heavy and superheavy nuclei should be carried out for a comprehensive analysis of the effects of different nuclear densities and NN potentials on the fusion dynamics.

\section*{Acknowledgements}
This work has been supported by the Science Engineering Research Board (SERB) File No. CRG/2021/001229, Sao Paulo Research Foundation (FAPESP) Grant 2017/05660-0, CNPq Grant 30313/2021-6, INCT-FNA Project 464898/2014-5, and FOSTECT Project No. FOSTECT.2019B.04. 


\end{document}